# REANN: A PyTorch-based End-to-End Multi-functional Deep Neural Network Package for Molecular, Reactive and Periodic Systems

Yaolong Zhang[*], Junfan Xia, and Bin Jiang[*]

*School of Chemistry and Materials Science, Department of Chemical Physics, Key Laboratory of Surface and Interface Chemistry and Energy Catalysis of Anhui Higher Education Institutes, University of Science and Technology of China, Hefei, Anhui 230026, China*

[*]: corresponding authors: ylzustc@mail.ustc.edu.cn; bjiangch@ustc.edu.cn

**Abstract**

In this work, we present a general purpose deep neural network package for representing energies, forces, dipole moments, and polarizabilities of atomistic systems. This so-called recursively embedded atom neural network model takes both advantages of the physically inspired atomic descriptor based neural networks and the message-passing based neural networks. Implemented in the PyTorch framework, the training process is parallelized on both CPU and GPU with high efficiency and low memory, in which all hyperparameters can be optimized automatically. We demonstrate the state-of-the-art accuracy, high efficiency, scalability, and universality of this package by learning not only energies (with or without forces), but also dipole moment vectors and polarizability tensors, in various molecular, reactive, and periodic systems. An interface between a trained model and LAMMPs is provided for large scale molecular dynamics simulations. We hope that this open-source toolbox will allow future method development and applications of machine learned potential energy surfaces and quantum-chemical properties of molecules, reactions, and materials.

## 1. Introduction

Atomistic simulations are powerful computational tools in understanding molecular spectroscopy, reaction dynamics, and energy/charge transfer processes in complex systems at the microscopic level[1]. Potential energy surfaces (PESs) and molecular electronic properties are prerequisite ingredients in such simulations. For small to medium-sized systems, these properties can be computed by ab initio calculations, with which molecular dynamics (MD) simulations can be performed on-the-fly. However, this so-called ab initio molecular dynamics (AIMD) approach is computationally expensive, limiting the size of investigated systems and the timescale of simulations.

In recent years, intense efforts have been devoted to developing various machine learning (ML) representations of PESs and quantum chemical properties based on ab initio data[2-15]. A recent special issue on Chemical Reviews has summarized the significant progress in developing ML representations for various purposes, including ground state and excited state PESs and electronic properties for small molecules, reactions, and materials, and so on[16-20]. These ML models for interaction potentials (or tensorial properties) differ mainly in the way of handling the intrinsic translational, rotational, and permutational invariance (or equivalence) symmetry of the system. For small molecular and reactive systems, it is natural to integrate (symmetrized) molecule-wise descriptors based on internal coordinates (or functions of interatomic distances), with linear regression (LR)[21, 22], spline interpolation[23], modified Shepard interpolation (MSI)[24], interpolating moving least-squares (IMLS)[25], and more recently neural networks (NNs)[5, 26-29] and kernel-based regression (KBR)[30-34], with different ways of symmetrization. Permutationally invariant polynomials (PIPs)[21] and fundamental invariants (FIs)[35] are classic examples of this kind of descriptors. Such approaches have been widely applied in small systems in gas phase and/or at gas-surface interface[36, 37]. However, their complexity increases dramatically with respect to the molecular size so that they are not suitable for molecules with more than ~15 atoms[38] and large periodic systems.

An alternative method designed for solving high-dimensional problems was first proposed by Behler and Parrinello[2] in their neural network approach (later referred to as BPNN). In this class of atomistic ML representations, the total energy is decomposed into atomic contributions dependent on their atom-center local environments. This concept has actually been rooted for long in many empirical force fields based on physically-motivated functions, *e.g.* the embedded atom method (EAM)[39]. What is new is that each atomic energy is now represented by an element-dependent ML model based on atomic descriptors[40] that encode the many-body correlations within an atomic local environment. This expression naturally satisfies the permutational invariance and enables a linear scaling with respect to the number of atoms in the system. In addition, once trained successfully, it can be in principle extended to describe larger systems consisting of the same type of atomic structures. The complexity is cast into the atomic descriptor and truncated by the cutoff radius. Since then, there have been numerous atomistic ML models differing in the physical concepts and/or mathematical expressions in their descriptors, taking advantage of NNs[7, 8, 10, 13, 41-45], KBR[3, 46-48], or LR[11, 12, 49-51], to just name a few. With few exceptions[7], these hyperparameters of atomic descriptors, if there is any, are often fixed during the training process, although they can be in principle optimized together with the ML model itself for a given training set. Various empirical or data-driven algorithms have been proposed to wisely select these atomic features to optimally describe atomic structures[8, 43, 52, 53]. This atomistic ML representation has been extended to preserve the proper rotational equivalence of tensorial molecular properties such as the dipole moment and polarizability and various models have been developed[54-60].

A more flexible way to construct an adequate description of an atomic structure is to recursively refining atomic features between every central atom and its neighbor atoms with respect to the reference data, through a message-passing type of deep neural network architecture[6, 9, 14, 60-66]. This so-called message-passing neural network (MPNN) representation allows the local environmental information of a central atom to be progressively updated by its connections with neighbors, and each neighbor's connections with neighbor's neighbors, and so on and so forth. This idea was inspired by the graph neural network (GNN)[67] model simplifying a molecule structure as a topologic graph, in which atoms are graphed as nodes that are interconnected by edges. Consequently, the first a few MPNN models, for example, SchNet and PhysNet were designed to only pass the edge (two-body) feature vector in an iterative way[6, 9]. More recently, incorporating the iterative update of angular features was found to improve the representability of the MPNN model[14, 60, 64-66, 68].

In practice, MPNN models learn a suitable representation of atomic structures and total energies concurrently in a data-driven manner[62]. They are thus usually recognized as end-to-end models, different from conventional atomic descriptor-based ML approaches with fixed hyperparameters. In this regard, the former are supposed to require less human intervention, while the latter can be more physically interpretable than the former[69]. Interestingly, our recent work[64] indicated that atomic descriptor-based NN models can be easily upgraded to MPNN ones when making the parameters associated with a neighbor atom dependent on its own local environment. In this spirit, we proposed a novel recursively embedded atom neural network (REANN) model[64] and showed that how the message-passing of three-body (or higher) features allows for a complete representation of the atomic structure without explicitly computing high-order correlations. It was also demonstrated that the message-passing form captures part of nonlocal interactions beyond the cutoff radius, allowing us to use a much smaller cutoff to achieve sufficient accuracy.

In company with these advances in the methodology of the ML representation, a diversity of open source packages have been developed for learning PESs and other properties[3, 9, 10, 12, 14, 24, 41, 48, 54, 60, 65, 66, 70-86]. These packages have lowered the technical barrier to ML applications in realistic systems and allowed for direct comparison among different ML models. In this work, we release an end-to-end implementation of REANN model taking advantage of the deep learning and automatic differentiation characteristics in PyTorch[87]. This implementation combines both benefits of descriptor-based and message-passing based representations, enabling optimizing all model parameters in the course of training. In addition, this package also allows for learning tensorial properties such as the dipole moment vector and polarizability tensor. We present comprehensive tests showing the state-of-the-art accuracy of the REANN model across bound molecules, reactive systems in gas phase and gas-surface interface, as well as periodic materials.

## 2. Methods

The original EANN model[8] is inspired by the EAM model, which takes the atomic decomposition form of total energy, namely $E=\sum_{i=1}^{N} E_i$ for an $N$-atom system, and expresses the atomic structure in terms of embedded atom densities (EADs). Each EAD descriptor is evaluated by the square of the linear combination of Gaussian-type atomic orbitals (GTOs) located at neighbor atoms associated with element-dependent coefficients. A group of EADs forms the input vector of the atomic NNs that output the corresponding atomic energy. On this basis, the more advanced REANN model makes each atomic orbital coefficient recursively adaptive to its atomic environment by an atomic NN in a similar way as the atomic energy, taking an essentially equivalent form of an MPNN model[64]. More details on the theory have been published before[8, 64, 88], along with some successfully applications[89-95]. We will hereafter focus here on the implementation of the REANN program on the PyTorch platform[87].

A schematic architecture of the REANN program is shown in Fig. 1. In general, Cartesian coordinates of neighbor atoms of each central atom are converted to a family of GTOs once only. Starting from the first iteration, these GTOs are repeatedly combined with recursively refined orbital coefficients to generate EADs, which are then passed to the EANN framework to predict orbital coefficients in the next iteration, as illustrated in Fig. 1(a). The maximum number of coefficient updates (or equivalently message-passings) is hereafter referred to as $T$. After $T$ iterations, the final EANN model will output the required property (energy or charge). In this scenario, REANN consists of ($T$+1) recursively connected EANN models. The dashed box in Fig. 1(a) means that this iterative procedure is optional and will be absent if $T$=0. The resultant REANN model will degrade to a conventional EANN model. Fig. 1(b) shows that an EANN iteration includes $N$ atomic density modules and $N$ atomic NN modules. Each NN module will contain several residual network blocks as shown in Fig. 1(c). The structure of a residual block is depicted in Fig. 1(d), consisting of multiple sub-blocks and a shortcut connection connecting the input and the output. Fig. 1(e) displays the layers of each sub-block including activation, normalization, dropout and linear mapping. Next, we will describe each module in detail.

## 2.1 Density modules

Each density module corresponds to an atomic environment of a central atom described by EADs. To calculate EADs, we first define a set of GTOs located at its neighbor atoms within a cutoff radius ($r_c$) as follows,

$$\varphi^m_{l_x l_y l_z}(\vec{\mathbf{r}}_{ij}) = (x_{ij})^{l_x}(y_{ij})^{l_y}(z_{ij})^{l_z} \exp\left[-\alpha_m(r_{ij}-r_m)^2\right] f_c(r_{ij})^l, \quad (1)$$

where $\hat{\mathbf{r}}_{ij} = \hat{\mathbf{r}}_j - \hat{\mathbf{r}}_i$, $x_{ij} = x_j - x_i$, $y_{ij} = y_j - y_i$, and $z_{ij} = z_j - z_i$ are the Cartesian coordinate vector and its three components of a neighbor atom $j$ relative to the corresponding central atom $i$, with $r_{ij}$ being their distance, $l=l_x+l_y+l_z$ specifies the orbital angular momentum (e.g., $l$=0 for the $s$ orbital, $l$=1 for the $p$ orbital, etc.), $\alpha_m$ and $r_m$ are hyperparameters to determine the center and the width of the radial Gaussian function, $f_c(r_{ij})$ is a cosine type cutoff function that makes the interaction smoothly decayed to zero at $r_c$ and is continuous up to its second-order derivative,

$$f_c(r_{ij}) = \begin{cases} [\cos(r_{ij}\pi/r_c)+1]^2, & 0 < r_{ij} \leq r_c \\ 0, & r_{ij} > r_c \end{cases}. \quad (2)$$

In earlier versions of EANN[8, 88], we simply take the linear combination of these primitive GTOs of neighbor atoms with the same shape ($l$, $\alpha_m$ and $r_m$) to form molecular orbitals in the local environment and then mimick an embedded electron density feature of the central atom by the square of a molecular orbital. Here, we find that it is beneficial to first mix atomic orbitals with different shapes and angular momenta before making the molecular orbital. This is indeed the ideas of hybridization and contraction in quantum chemistry for obtaining a more reasonable atomic basis set[96]. In this way, the $n$th EAD feature of the central atom $i$ will be expressed as,

$$\rho_i^n = \sum_{l=0}^{L} \sum_{l_x,l_y,l_z}^{l_x+l_y+l_z=l} \frac{l!}{l_x!l_y!l_z!} \left[ \sum_{j \neq i}^{N_c} c_j \sum_{m=1}^{n_{wave}} d_m^n \varphi_{l_x l_y l_z}^m (\hat{\mathbf{r}}_{ij}) \right]^2. \quad (3)$$

Here, $L$ is the maximum orbital angular momentum of primitive GTOs, $d_m^n$ is contraction coefficient of the $m$th primitive GTO to generate the $n$th contracted GTO, $c_j$ represents the orbital coefficient of the $j$th neighbor atom that is element-dependent for the first iteration of EANN, $n_{wave}$ is the number of primitive GTOs for a given $l$ and $N_c$ is the number of neighbor atoms. Interestingly, the linear combination of $n_{wave}$ primitive GTOs is something like the construction of an STO-nG atomic basis set, the linear combination of the contracted GTOs over $j$ and $l$ gives rise to a hybrid molecular orbital. In practice, we reorder the summation over $c_j$ and $d_m^n$ in evaluating Eq. (3) to achieve much higher efficiency because $d_m^n$ is unchanged in different neighbor atoms. The advantage of Eq. (3) is more easily seen by explicitly expanding the square term. According to the multinomial theorem, we can rewrite the Eq. (3) in the following form[49],

$$\rho_i^n = \sum_{l=0}^{L} \sum_{l_x,l_y,l_z}^{l_x+l_y+l_z=l} \frac{l!}{l_x!l_y!l_z!} \sum_{j,k \neq i}^{N_c} c_j c_k [f_c(r_{ij})f_c(r_{ik})]^l (x_{ij})^{l_x}(y_{ij})^{l_y}(z_{ij})^{l_z}(x_{ik})^{l_x}(y_{ik})^{l_y}(z_{ik})^{l_z} \\ \times \sum_{m,m'=1}^{n_{wave}} d_m^n d_{m'}^n \exp\left[-\alpha_m(r_{ij}-r_m)^2\right]\exp\left[-\alpha_{m'}(r_{ik}-r_{m'})^2\right] \quad (4)$$

Here $n_{wave}$ can be viewed as the "n" label in a STO-nG basis and $\rho_i^n$ actually contains information of $n_{wave}(n_{wave}+1)/2$ different combinations of primitive GTOs for each $l$. In other words, this scheme uses only $n_{wave}$ primitive GTOs to generate atomic features that have the same spatial resolution as using $n_{wave}(n_{wave}+1)(L+1)/2$ GTOs in the original version of EANN[8], which greatly reduces the computational cost while retaining the representability.

### 2.2 NN modules

In the NN modules, these EADs generated in the density modules will be input to the corresponding atomic NNs to compute atomic orbital coefficients for the iteration number $t < T$,

$$c_j^t = g_j^{t-1}\left(\rho_j^{t-1}\left(\mathbf{c}_j^{t-1}, \mathbf{r}_j^{t-1}\right)\right), \quad (5)$$

or atomic energies (or charges, see below) when $t = T$,

$$E_j = g_j^{T-1}\left(\rho_j^{T-1}\left(\mathbf{c}_j^{T-1}, \mathbf{r}_j^{T-1}\right)\right), \quad (6)$$

where $g_j^{t-1}$ denotes the $j$th atomic NN module in the $t$th iteration. As illustrated in Fig. 1(c), each NN module contains several residual blocks[97] to avoid the problem of vanishing gradients while training deep NNs using gradient-based optimization method.

Each residual block consists of multiple sub-blocks, each involving a nonlinear activation layer, a normalization layer, a dropout layer, and a linear layer, respectively, as displayed in Fig. 1(e). In the activation layer, two different activation functions as follows can be used[68],

$$\sigma_1 = \eta x/[1+\exp(-\beta x)] \quad \text{or} \quad \sigma_2 = \eta x / \sqrt{1+(\beta x)^2}. \tag{7}$$

The former has a similar shape as the well-known Relu function and the latter is a tangent hyperbolic like function. The behaviors of these two activation functions vary with learnable parameters $\eta$ and $\beta$. Note that $\sigma_1$ will approach the Relu function as $\beta$ increases but with a slope of $\eta$ instead of 1, and will degrade to a linear function ($y=\eta x/2$) for a tiny $\beta$. On the other hand, $\eta$ and $\beta$ in $\sigma_2$ control the function values that reach saturation and the effective range of nonvanishing gradients respectively. A layer normalization procedure[98] is applied to normalize over all the hidden neurons in the same layer instead of the mini-batch during training, which is independent of the batch size and more suitable for the database containing individual molecules. A dropout layer[99] is adopted to randomly zero some neurons with a probability $p$ obtained from the Bernoulli distribution for regularization and prevent the co-adaptation of neurons. This is advantageous for improving the generalizability of NNs. Both Normalization and Dropout layers are optional. A linear layer will generate the input for the next sub-block.

Specifically, an EAD input vector first goes through a standard linear layer before entering these residual blocks, and the input of the previous residual network will be directly connected with the output to generate the input of the next one, as shown in Fig. 1(d). The output of the last residual black will be converted to an atomic energy (or charge) or atomic orbital coefficient by a nonlinear activation function and a linear layer (see Fig. 1(c)). Any derivatives (*e.g.* atomic forces) can be derived analytically from the output back to atomic Cartesian coordinates through the autograd mechanism implemented in PyTorch.

**2.3 Prediction of tensorial properties:**

In addition to learning energies and forces, this REANN package allows one to learn dipole moments and polarizabilities based on the theory published elsewhere[59]. The basic atomistic framework of learning these tensorial properties is similar to that of learning potential energies with respect to translation and permutation symmetry. However, the former requires additional treatment to preserve the equivariant property with respect to rotation. For example, one can simply replace the atomic energy with the atomic charge as the output of the final EANN model and use the sum of its product with the atomic Cartesian coordinate vector to express the permanent dipole moment vector ($\vec{\mu}$) that transforms correspondingly as the system rotates, as shown in Fig. 1(a)

$$\vec{\mu} = \sum_{i=1}^{N} q_i \vec{r}_i, \qquad (8)$$

where $q_i$ is the atomic charge of $i$th atom and $\vec{r}_i = (x_i, y_i, z_i)^T$ is the Cartesian coordinate vector. This strategy was first proposed by Marquetand and coworkers[55] and has been made by default together with learning energies and forces in some NN packages, such as TensorMol[74] and PhysNet[9]. Note that the dipole moment can be also expressed in a symmetric yet more complex way in terms of internal coordinate based molecule-wise descriptors, with the global molecular descriptor, *e.g.* PIPs[100].

On the other hand, molecular polarizability (**α**) is a symmetric second-rank tensor in 3×3 dimension. Indeed, we make use of the REANN architecture to construct three 3×3 matrices with the same rotational equivariance as **α** and their sum properly fulfills the symmetry requirement of **α**. The first matrix is the multiplication of a first derivative matrix (3×*N*) of a latent scalar output of REANN model (say, a total energy like quantity, *E'*) with respect to atomic coordinates with its transpose (*N*×3), namely,

$$\boldsymbol{\alpha}^{NN1} = \left(\partial E'/\partial \mathbf{r}\right)\left(\partial E'/\partial \mathbf{r}\right)^T, \tag{9}$$

where **r** is the 3×*N* Cartesian coordinate matrix of the system. However, $\boldsymbol{\alpha}^{NN1}$ is a semidefinite matrix by construction, while the target molecular polarizability is not necessarily so. The second matrix $\boldsymbol{\alpha}^{NN2}$ is therefore defined to remove this restriction by the following way,

$$\boldsymbol{\alpha}^{NN2} = \mathbf{r}(\partial E'/\partial \mathbf{r})^T + \left(\partial E'/\partial \mathbf{r}\right)\mathbf{r}^T. \tag{10}$$

Moreover, it is important to realize that both $\boldsymbol{\alpha}^{NN1}$ and $\boldsymbol{\alpha}^{NN2}$ will become a rank-deficient matrix for a planar geometry, while the molecular polarizability tensor is not. The third scalar matrix $\boldsymbol{\alpha}^{NN3}$ is introduced to correct this, whose identical diagonal elements can be learned from a scalar virtual output of a separate REANN model. Summing the three matrices together will guarantee the full symmetry of **α**,

$$\boldsymbol{\alpha} = \boldsymbol{\alpha}^{NN1} + \boldsymbol{\alpha}^{NN2} + \boldsymbol{\alpha}^{NN3}. \tag{11}$$

**2.4 Training and hyperparameters**

Our REANN program implemented in the PyTorch framework can be executed with both GPU and CPU. Taking advantage of the autograd mechanism and Distributed DataParallel features in PyTorch, the training process is highly parallelized. In Fig. 2, we summarize the learning workflow of the REANN package. The training process can be divided into four parts: information loading, initialization, dataloader and optimization. First, the dataset containing the system information and input files ("input_nn" and "input_density") containing the model information are loaded in the module "src.read". In this module, the default distributed process group will be initialized for distributed training. Second, the "run.train" module utilizes the loaded information to initialize various classes, including property calculator, dataloader, and optimizer, and to build connections between each process. For each process, an additional thread will be activated in the "src.dataloader" module to prefetch data from CPU to GPU in an asynchronous manner. Meanwhile, the optimization will be activated in the "src.optimize" module once the first set of data is transferred to the GPU. During optimization, all NN parameters and hyperparameters are optimized using the mini-batch gradient descent technique with the AdamW optimizer[101] through automatic differentiation. The loss functions of the energy (and force whenever available) polarizability are defined as follows, respectively,

$$\mathcal{L}(\mathbf{w}) = \sum_{m=1}^{N_b}[\lambda_E \times (E_m^{NN} - E_m^{Ref})^2 + \lambda_F \times |\mathbf{F}_m^{NN} - \mathbf{F}_m^{Ref}|^2]/N_b , \qquad (12)$$

$$\mathcal{L}(\mathbf{w}) = \sum_{m=1}^{N_b}[|\mathbf{\mu}_m^{NN} - \mathbf{\mu}_m^{Ref}|^2]/N_b , \qquad (13)$$

$$\mathcal{L}(\mathbf{w}) = \sum_{m=1}^{N_b}[|\mathbf{\alpha}_m^{NN} - \mathbf{\alpha}_m^{Ref}|^2]/N_b , \qquad (14)$$

and minimized separately. Here, $N_b$ is the size of the mini-batch and the superscripts of each quantity refer to the NN predicted and reference quantities, respectively, $\lambda_E$ and $\lambda_F$ represent the weights of the energy and force in the loss function used in the construction of PES.

In a practical training process for the models reported in this work, the training set will be shuffled for each epoch and an exponential moving average of all model parameters is imposed using a smoothing factor of 0.999. Then the model with the averaged parameters is evaluated on the validation set. The learning rate with an initial value of 0.001 will be decayed by a factor of 0.5, whenever the validation loss does not decrease for consecutive "patience_epoch" epochs, typically 100~300 dependent on the batchsize and the size of the training data. Training is stopped when the learning rate drops below $1\times10^{-5}$ and the model that performs best on the validation set is saved for further investigation.

In our previous work[64], the minimum number of iteration ($T_{\min}$) that warrants a complete description is estimated by $T_{\min} = \left[\log_2(N_c(N_c-1)/6+1)\right]-1$, where [] rounds up the value to its nearest integer. For molecular systems in this work, $r_c$ is sufficiently large to include all atoms in the system and $T$ is equal to $T_{min}$ ($N_c=N$), while $r_c$ is truncated until the fitting errors do not significantly change and $T$ is set to 3 for periodic systems to maintain the balance of accuracy and efficiency. The NN structures for those systems are given in Table I. These settings are expected to be suitable for most systems of interest.

A trained model can be compiled with the Torch JIT embedded in PyTorch to create a serialized and optimized file, which can be later loaded and executed in C++. We have provided an interface to LAMMPS[102] by creating a new pair_style invoking this representation for highly efficient MD simulations[103].

## 3. Result and discussion

In this section, we validate the performance of the REANN approach in a variety of systems ranging from molecules with small and medium sizes, reactive systems including both gas phase and gas-surface reactions, to periodic systems with hundreds of degrees of freedom. We choose to refit these publicly available datasets so that the new REANN model can be compared with previously published ones. All datasets can be found in original publications unless stated elsewhere.

### 3.1 Molecules

### 3.1.1. MD17

MD17 is a collection of structures, energies, and forces from AIMD simulations of several organic molecules computed by density functional theory (DFT) and has become a testing ground for new ML models[6, 8-10, 32, 104, 105]. In Tables II and III, we compare the test mean-average-errors (MAEs) over the entire dataset predicted by potentials trained with 1000 (1K) and 50000 (50K) randomly selected configurations by the REANN and several other well-established methods, respectively. An independent model is trained for each molecule and the best result for each molecule is marked in bold. With only 1K data points, the REANN potentials are much more accurate than other NN-based ones, reaching comparable or even better accuracy as the symmetric Gradient Domain Machine Learning (sGDML)[32] and FCHL[105] models that are based on kernel regression and commonly recognized to be more data-efficient than NN-based alternatives[18]. The superior data-efficiency of the REANN approach is more explicitly seen in Fig. 3, where MAEs of sGDML and REANN potentials are compared as a function of the number of training points from 200 to 1000. The prediction errors of both methods quickly decrease with the increasing size of training set. Even with as few as 200 points, REANN performs comparably well or better than sGDML for most molecules. On the other hand, with 50K data points, where the sGDML approach is not feasible because of its interpolation nature, the REANN model performs better than those atomic descriptor-based deep NN models, *e.g.* deep potential MD (DPMD)[104] and Gaussian-moment (GM) neural network[10] (GM-sNN and GM-dNN). Our REANN model is also more accurate than SchNet[6] and as accurate as PhysNet[9], both are models of the MPNN type. In any case, the REANN potentials are much more accurate than the original EANN ones[8]. This is due on one hand to the full optimization of hyperparameters in the new implementation and on the other hand to the iterative update of the orbital coefficients, although the performance of the original EANN model can be improved by selecting more representable EADs[53]. These results indicate the state-of-the-art accuracy of this REANN package in both data-insufficient and data-sufficient regimes.

### 3.1.2. CH$_4$

The CH$_4$ dataset has been designed by Ceriotti and coworkers[106, 107] at the DFT level to test the representability and completeness of an atomic descriptor. This dataset includes ~7.7 million configurations with H atoms randomly placed in a 3 Å sphere of the C center while excluding structures with any two H-H distances less than 0.5 Å. Consequently, there exist many hardly-distinguishable and unphysical configurations in this dataset, with an energy range up to 70 eV. In Fig. 4, we compare the test root-mean-square-errors (RMSEs) of various models over 80K randomly chosen configurations out of the whole set, as a function of the number of randomly chosen training data points. In an earlier work[64], we have demonstrated that MPNN models[6, 65, 66, 68], thanks to their implicit incorporation of more complete many-body correlations, are generally more accurate than these with locally truncated many-body descriptors, *e.g.* three-body + four-body (3B+4B) [106] or contracted five-body (5B)[107] based NN models. Here, the present REANN model leads to systematically smaller RMSEs than all previous models including the earlier REANN version. This is due largely to the new implementation enabling optimizable activation functions and layer normalization, as well as the improved EAD expression. This new REANN model shows an extraordinary data efficiency and achieves comparable RMSEs with about two orders of magnitude fewer data points than the (3B+4B) feature based NN representation.

**3.2. Reactive systems**

**3.2.1 $N_4$**

Reactive PESs, in particular those for describing reactive scattering, are typically more difficult to learn, as they involve broader configuration spaces including asymptotic region, heavily distorted configurations, as well as the breakage/formation of chemical bonds. Here, we test the performance of the REANN method against two representative systems. The first system is $N_4$, whose energies have been obtained by multi-reference or single-reference methods in separate regions. This $N_4$ dataset includes 21406 data points (energies only), covering a huge configurational space and multiple complicated reaction channels (e.g. $N_2+N_2$, $N_2+N+N$, $N_3+N$ and $4N$) with an energy range of ~87 eV. More details on the dataset can be found elsewhere[108-110]. In Table IV, the trained REANN potential for $N_4$ using this data set is compared with previous ones using different methods[108-110] with respect to the fitting errors. The fitting RMSE of the REANN potential over all data points is 65 meV, which is merely about half of that of the previously best MB (many-body)-PIP-NN potential[110] (~120 meV). This result suggests that the REANN method is at least as accurate as the PIP-NN approach, one of the most accurate ML approaches in fitting gas phase PESs[36].

### 3.2.2 $CO_2$+Ni(100)

The second reactive system is a gas-surface reaction, namely the dissociative chemisorption of $CO_2$ on a movable Ni(100) surface[111]. This $CO_2$+Ni(100) system includes not only the reactive DOFs of the $CO_2$ molecule but also the periodic DOFs of the surface. The whole dataset contains ~10000 points extracted from 2200 AIMD trajectories, with their energies and forces computed by plane-wave DFT. The Ni(100) surface was modeled by a four layer slab in a $3 \times 3$ unit cell with the top three layers relaxed, yielding 39 atoms (90 moveable DOFs) in a periodic system. More details about the generation of this dataset can be found in previous publications[111, 112]. This dataset has been fitted to a BPNN PES[111], with its prediction RMSEs of all data points being 15.1 meV for energies and 26.1 meV/Å for forces, respectively. As seen in Table IV, the present REANN fit with the same dataset nearly half of the prediction errors for both energies (8.5 meV) and forces (12.2 meV/ Å), compared with the previous BPNN PES[111]. Here, the significant improvement of the current REANN potential over the earlier BPNN one is likely a combination effect of the optimization of hyperparameters and the better description of atomic structures in the former model. In addition to the RMSEs, we have calculated the phonon spectrum of the Ni(100) surface represented by a four-layer slab in a 3×3 cell. As shown in Fig. 5, REANN reproduces the DFT-derived phonon band structure as well as the density of phonon states of Ni(100), which validates the reliability of this PES in describing surface vibration.

### 3.3. Periodic systems:

### 3.3.1 bulk water

Bulk water is an important condensed phase system that has been frequently used to demonstrate the power of various ML methods[7, 113]. We choose to fit the ab initio data reported by Cheng et al.[113], which consists of energies and forces for 1593 diverse structures containing 64 water molecules in a periodic box computed by DFT. These data have been originally used to construct a BPNN potential[113] with the test RMSEs of 2.3 meV/atom for energies and 120 meV/Å for forces, and later a hyperparameter-fixed EANN potential with the test RMSEs of 2.1 meV/atom for energies and 129 meV/Å for forces[88]. In Fig. 6, we compare the REANN potentials with these previous ones with respect to the test RMSEs as a function of the cutoff radius. It is interesting to note that the REANN model (and other MPNN models as well) can progressively pass neighbor's correlations beyond the cutoff sphere to incorporate some nonlocal effects. This feature leads the REANN potential to a prediction error as low as the previous BPNN and EANN models, even with a cutoff radius as small as 3.0 Å. On the other hand, the reliability of the REANN model quickly increases with increasing $r_c$, as more atoms are included in the local environment and higher-order interatomic correlations are more important which can be explicitly involved by message-passing. This feature makes the REANN representation in principle more accurate than local descriptor-based counterparts. Remarkably, the REANN model leads to more or less converged prediction RMSEs of 0.8 meV/atom (energy) and 47 meV/Å (force) at $r_c$ = 5.5 Å, about 1/3 of those of BPNN and EANN potentials. This result is consistent with our earlier REANN result using an old version of EADs[64]. To check the robustness of the REANN ($r_c$=5.5 Å), we have performed classical MD simulations for liquid water containing 64 $H_2O$ molecules in a periodic box at 300 K for 20 ps with a time step of 1 fs. To avoid the influence of quantum effects, for our purpose, we choose to compare in Fig. 7 the calculated O-O radial distribution function (RDF) with the experimental data[114]. The excellent agreement between theory and experiment indicate this REANN water potential is free of unphysical features and robust for MD simulations.

**3.3.2 bulk SiO$_2$**

The $SiO_2$ dataset was generated by Yanxon et. al.[80] who extracted configurations in AIMD trajectories at different temperatures, including liquid, amorphous and various crystallines (α-quartz, α-cristobalite, and tridymite). The whole dataset contains 10000 configurations with energies and atomic forces calculated at DFT/PBE level. 1316 configurations are randomly selected for training and the test RMSE is evaluated on the rest of the dataset. As displayed in Table V, the REANN model leads to much smaller errors for energies and forces than various models reported in the original work[80]. Note that Yanxon et. al have indeed implemented EAD as one of the descriptors in their PyXtal_FF library[80], which shows a superior efficiency yet a worse accuracy of EANN compared with other ML models. When using a similar number of EAD features and NN parameters, our new EANN result (namely, REANN ($T$=0)) becomes more accurate than all three-body descriptor-based NN models in Ref. 80. The different performance of the two EANN implementations is likely due to the fact that the orbital coefficients were fixed in PyXtal_FF during training[80].

**3.4 Dipole moment vector and polarizability tensor**

After demonstrating the performance of the REANN program in learning energies and forces, we now apply it to learn dipole moments and polarizabilities of a number of molecular and periodic systems, including a water molecule, water-dimer, $(H_5O_2)^+$, and liquid water from datasets reported by Ceriotti and coworkers[56]. Ab initial calculations for $H_2O$, $(H_2O)_2$, and $H_5O_2^+$ were performed at the CCSD/d-aug-cc-pvtz level. Liquid water was described by a 32-water box and calculated at the DFT/PBE-USPP level, using a plane-wave basis set with a kinetic energy cutoff of 55 Ry[56]. Details on these datasets are referred to as original publications[56]. To compare with the results obtained by the symmetry-adapted Gaussian approximation potential (SA-GAP) method, half of 1000 configurations were randomly chosen for training and the rest half for testing, and the RMSEs relative to intrinsic standard deviation (r-RMSEs) of the testing samples are calculated. Table VI shows the SA-GAP and REANN results, along with those previously obtained from the hyperparameters-fixed EANN model. In spite of such a small amount of data, the EANN model outperforms the kernel-based SA-GAP model in most cases except for the dipole moment of the water dimer. The REANN approach further improves the representation of this system and $(H_5O_2)^+$, and lowers the fitting errors. As an example, we have performed classical MD simulations for the $H_2O$ molecule at 300K with the REANN learned potential and dipole models, removing the overall translation and rotation at each step. By Fourier transformation of the dipole derivative correlation functions, the IR spectrum of $H_2O$ has been obtained and presented in Fig. 8. There are three peaks at 1643, 3795, and 3929 cm$^{-1}$, corresponding to the bending, symmetric and anti-symmetric stretching modes, respectively. These values are somewhat larger than experimental ones, due presumably to the intrinsic error of ab initio data and/or the neglect of quantum effects. These results are sufficient to support that the REANN learned potentials and electronic properties can be used for spectroscopic simulations.

### 3.4 Efficiency and Scalability

It is important to demonstrate the efficiency and scalability of the REANN program. To this end, we take the liquid water system as an example and evaluate the computational cost of each MD step (*i.e.* per force evaluation) during NVT simulations performed with a varying number of water molecules in a cubic box under periodic conditions at a temperature of 300 K using LAMMPS. Fig. 9 compares the efficiency of using a single CPU core (Intel(R) Xeon 6132 2.60 GHz), where REANN scale linearly with respect to the total number of atoms in the system. This is an intrinsic feature of the atomistic type of ML models. Our REANN potential with $r_c$=5.5 Å is slightly less efficient than the previous BPNN potential, as a price of much higher accuracy. When $r_c$ decreases to 3.0 Å, the REANN potential reaches a similar accuracy to the BPNN potential at the speed of (~$1.2\times10^{-4}$ s/atom/step/CPU).

GPU can significantly accelerate the evaluation of the REANN model when the system size is large. Fig. 10 displays the computational cost of the REANN water potential using a single GPU (NVIDIA V100s). It should be noted that the acceleration ratio of GPU versus CPU increases with both the total number of atoms and the number of neighbor atoms within $r_c$. This is because GPU has thousands of inner cores and will enable the highest efficiency if these cores are fully loaded. A maximum acceleration ratio is about two orders of magnitude, which is reached at ~$10^4$ atoms when $r_c$=5.5 Å and ~$10^5$ atoms when $r_c$=3.0 Å at the speed of ~$1.4\times10^{-6}$ s/atom/step/GPU. Fig. 10 also shows that the GPU implementation of the REANN model is memory-efficient. For example, a single V100s having a 32 GB memory makes a MD simulation of a maximum of $5.3\times10^5$ atoms possible using the REANN potential with $r_c$=3.0 Å. This favorable memory requirement will be useful for scalable atomistic simulations.

Because of the correlation between different local environments during the iterative update of orbital coefficients, the multi-process parallelization algorithm of the REANN model is not as direct as a traditional descriptor-based NN model, in which the atomic energy is only dependent on the corresponding neighbor list of central atom. Instead, we test the parallel efficiency of the regular EANN model, namely REANN ($T$=0), which can simply take advantage of the parallel mechanism of atomistic potentials implemented in LAMMPS. Using bulk water as an example, the computational costs of evaluating the REANN($T$=0) potential for a fixed number of water molecules, as a function of the number of CPU cores and GPUs are presented in Fig. 11(a) and Fig. 11(b), respectively. Note that the number of water molecules needs to be large enough to maximize the computational power of GPU. These results clearly indicate that the REANN package can be almost linearly accelerated by CPU/GPU and the parallel efficiency of is over 90% in either case.

As we have stated previously[8, 88], the computational cost of an atomistic NN model is often dominated by the descriptor calculation, which mainly depends on $N_c$ and $n_{wave}$ in the REANN framework. Fig. 12(a) shows that the computational cost scales linearly with $N_c$ in the bulk water system. Fig. 12(b) to show that the computational cost does not increase significantly with $n_{wave}$, for example from $n_{wave}=4$ to $n_{wave}=32$, the computational time only increase about 50%. These favorable scalings are the advantages of the EAD descriptors by design.

The training cost of this REANN package is system dependent. For example, using a single NVIDIA V100S GPU, training the REANN potentials of molecules in the MD17 dataset with 50000 structures (energies and forces) typically took 3 to 9 days, because the RMSE converges very slowly to reach a very high accuracy of each model. Training the reported $CO_2$+Ni(100) potential took ~20 hours. Training these bulk water potentials in this work took 4 to 9 hours depending on the $n_{wave}$ and $r_c$.

**3.5 Comparison with other Software Packages**

There exist a number software packages all based on atomistic machine learning but with different features and superiorities. For instance, DPMD-kit[70] is well-known for its abundant and efficient interfaces with many popular MD simulation packages, enabling very large scale MD simulations for condensed phase systems. In comparison,

PhyNet[9] and SchNet[75] are more focused on molecular systems. Thanks to their message-passing nature, they have a larger potential to achieve better accuracy. The former additionally incorporates the long-range Coulomb interactions taking advantage of concurrently learned atomic charges. On the other hand, the kernel-based GAP package[3] is expected to suffer from the poor scaling with respect to the size of the dataset. However, it is considered to be more data-efficient than NN-based models. Here, our REANN package uses a message-passing type of NN model based on intrinsically efficient three-body EAD features, which is beneficial for constructing a complete representation of more complex atomic structures of both molecular and condensed phase systems. It is highly parallelized and memory efficient as powered by PyTorch.

**Conclusion**

To summarize, REANN is a multi-functional deep learning package based on PyTorch, which can be used to learn energies, forces, dipole moment vectors and polarizability tensors of atomistic systems. This program is highly modular and the training process can be parallelized with multiple GPUs or CPUs using the Distributed DataParallel. It has been extensively tested in various benchmark systems, validating its state-of-the-art accuracy, high efficiency, linear scalability, and good generalizability. By learning dipole moments, atomic charges are also obtained, which may be used for describing long-range charge-dipole interactions in the future. We note that the inference speed of a trained REANN model still needs improvement due to the intrinsic low efficiency in Python. Further acceleration may be achieved by implementing the trained/evaluated model using LibTorch in C++. In addition, since REANN and other MPNN models break down the local approximation and all atoms involved in the message passing are correlated, the multi-process parallelization algorithm for evaluation is not as straightforward and efficient as in classical force fields or conventional locally atomic descriptor-based ML models. This needs to be further optimized when performing large scale molecular dynamics simulations. Work along this direction is underway in our lab.

**Data Availability:**

The dataset of the $CO_2$+Ni(100) reactive system and the REANN package are openly available in github repository REANN[115]. A user's manual and several examples can be found there. Other datasets can be found in the corresponding original publications.

**Acknowledgement:** This work is supported by National Key R&D Program of China (2017YFA0303500), National Natural Science Foundation of China (22073089 and 22033007), Anhui Initiative in Quantum Information Technologies (AHY090200), and The Fundamental Research Funds for Central Universities (WK2060000017). Calculations have been done on the Supercomputing Center of USTC and Hefei Advanced Computing Center.

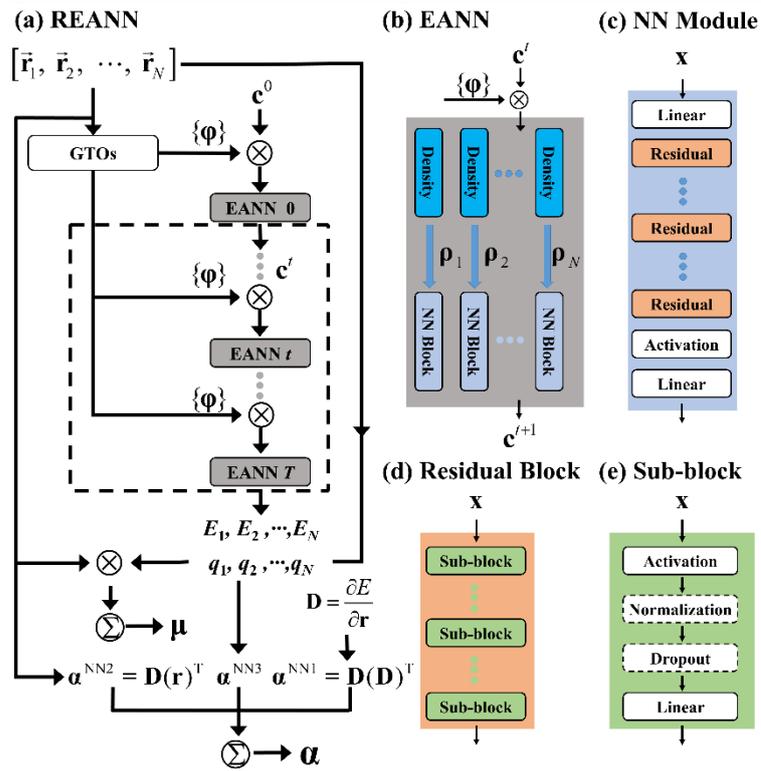

Fig. 1. Schematic decomposition and illustration of the REANN architecture.

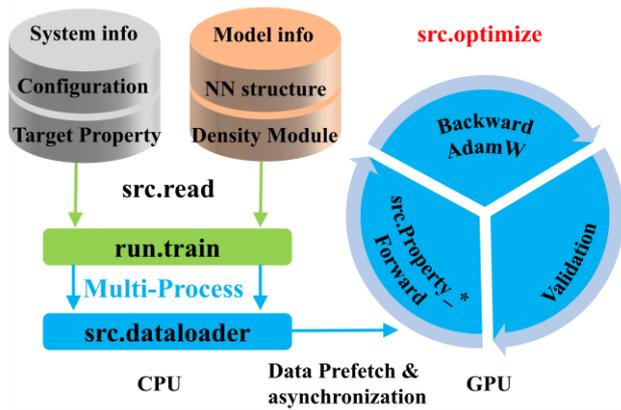

Fig. 2. Schematic workflow of the REANN training process. The same color used by "src.dataloader" and "src.optimize" indicates that they are asynchronously executed on CPU and GPU, respectively.

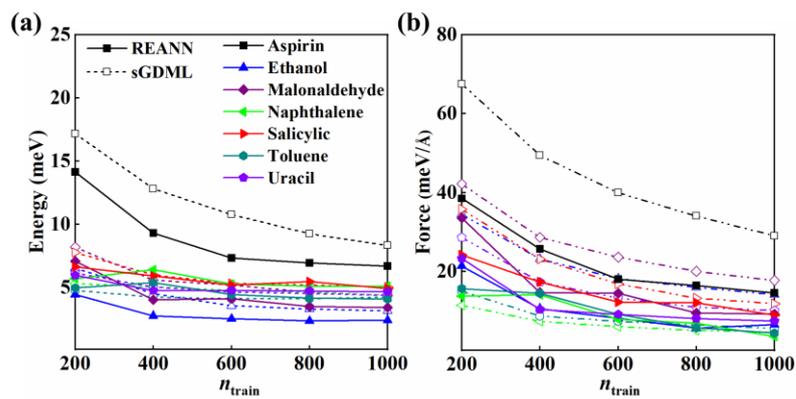

Fig. 3. Comparison of learning curves for energies (a) and atomic forces (b) predicted by sGDML and REANN potentials for several molecules in the MD17 dataset as a function of training points.

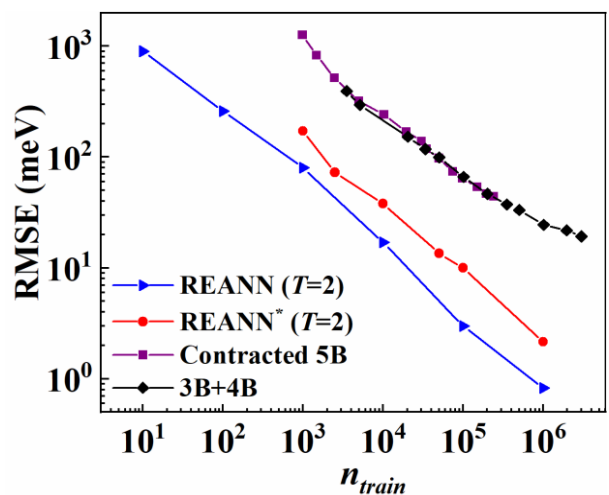

Fig. 4. Comparison of the test RMSEs of $CH_4$ energies of a random dataset as a function of training data ($n_{train}$) using the present REANN model (blue triangles), an earlier REANN model (red circles)[64] and two descriptor-based neural network models reported in Refs. [107] (Contracted 5B features, purple squares) and [106] (3B+4B features, black diamonds), respectively.

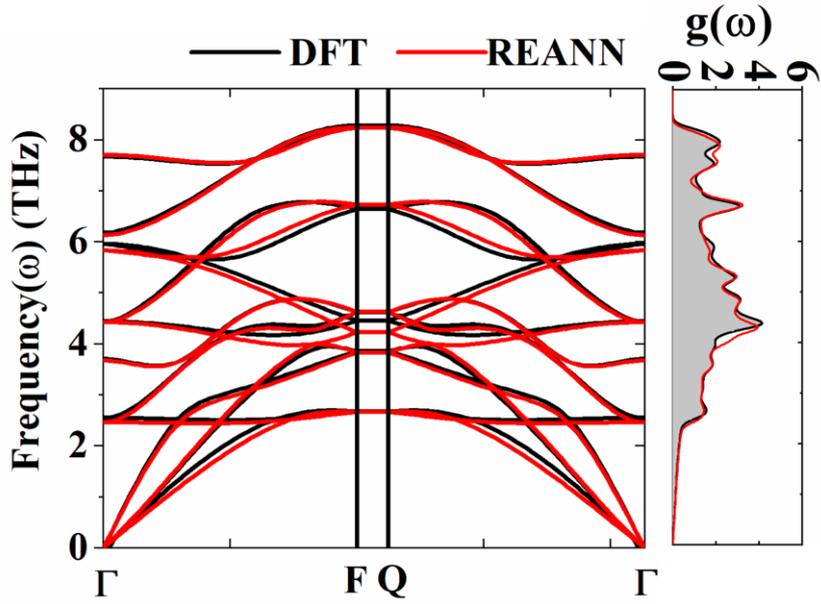

Fig. 5. Phonon dispersion of Ni(100) obtained by DFT and the REANN Pes for $CO_2$+Ni(100) along the path through the surface Brillouin zone given by the high-symmetry points Γ-F-Q-Γ (left) and corresponding phonon density of states as a function of the frequency of phonon modes (right).

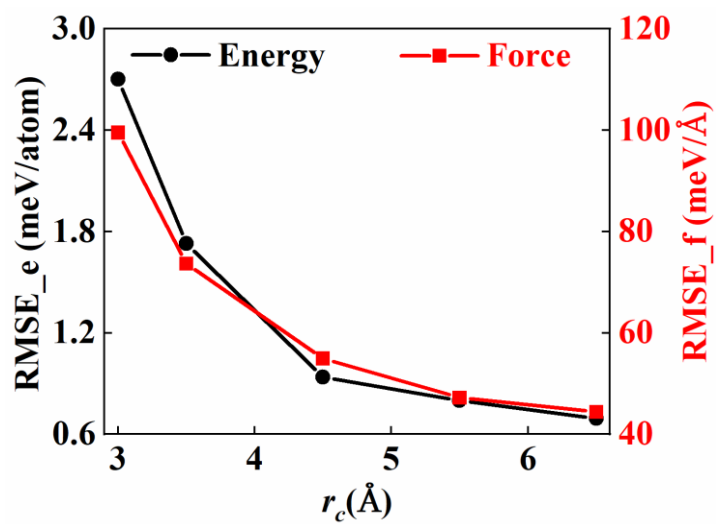

Fig. 6. Variation of the RMSEs of energies per atom (black circles, left tick labels) and atomic forces (red squares, right tick labels) of the liquid water dataset from Ref. [113], predicted by REANN with respect to the cutoff radius.

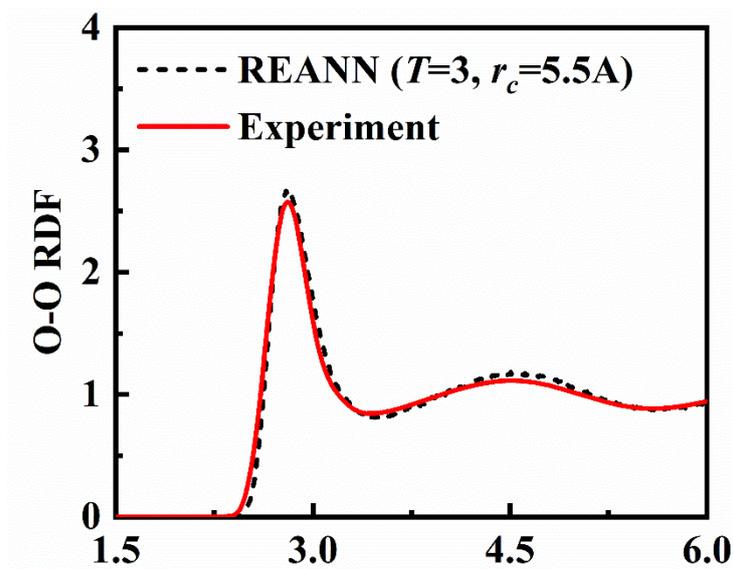

Fig. 7. Comparison of experimental[114] and theoretical O-O radial distribution functions of liquid water calculated by classical MD simulations using the REANN potential at 300 K.

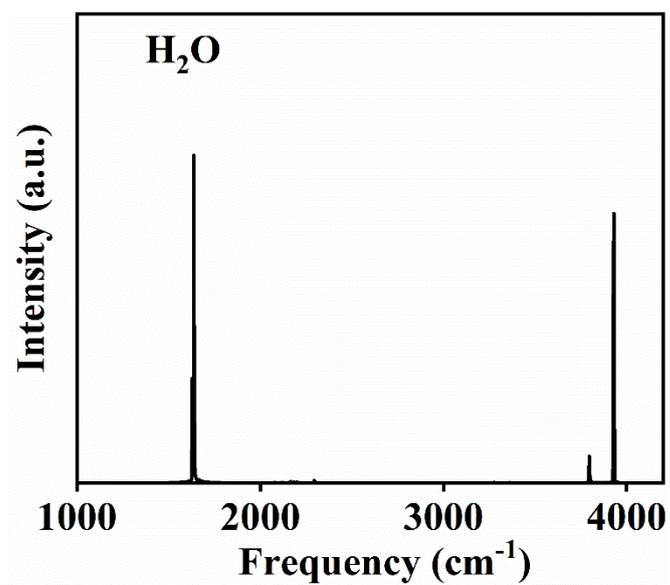

Fig. 8. The IR spectrum of $H_2O$ at 300 K calculated by classical MD simulations based on the REANN learned PES and dipole moment surface.

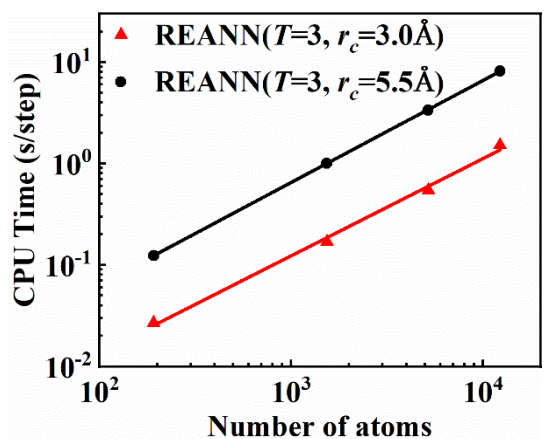

Fig. 9. Computational costs (CPU time per MD step) of the REANN potentials with different cutoff radius, as a function of total atoms in simulations of the liquid water (300 K), using a single core of the Intel(R) Xeon 6132 2.60 GHz processor.

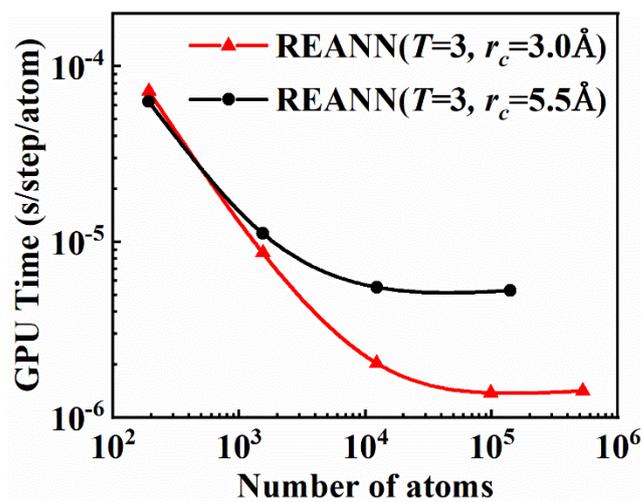

Fig. 10. Computational costs (GPU time per MD step per atom) of the REANN potentials with $r_c$=3.0 Å (black circles) and $r_c$=5.5 Å (red triangles), as a function of total atoms in simulations of the liquid water (300 K), using a single GPU (NVIDIA V100s).

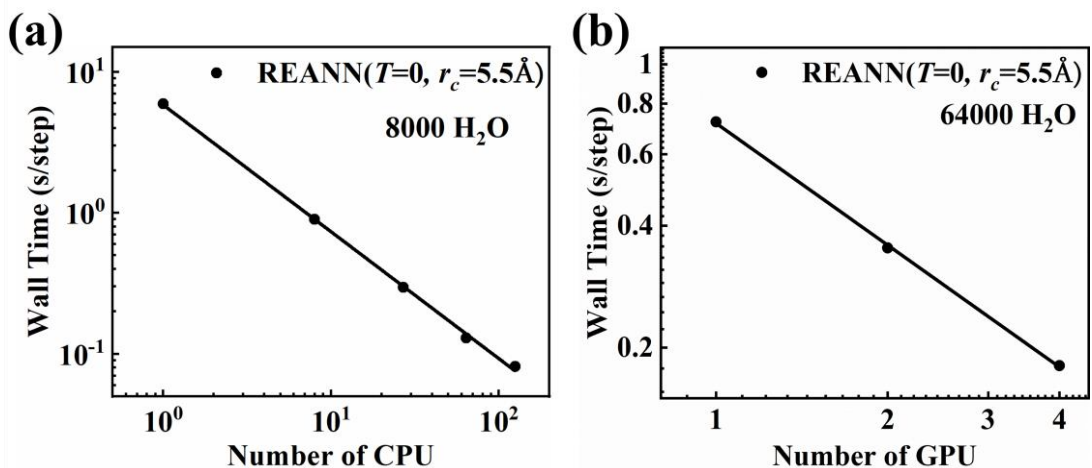

Fig. 11. (a) Computational costs (wall time per MD step) of the REANN potentials as a function of the number of (a) CPU (Intel(R) Xeon 6132 2.60 GHz), via classical MD simulations of liquid water (300 K) including 8000 $H_2O$ molecules and (b) GPU (NVIDIA V100s), via classical MD simulations of liquid water (300 K) including 64000 $H_2O$ molecules.

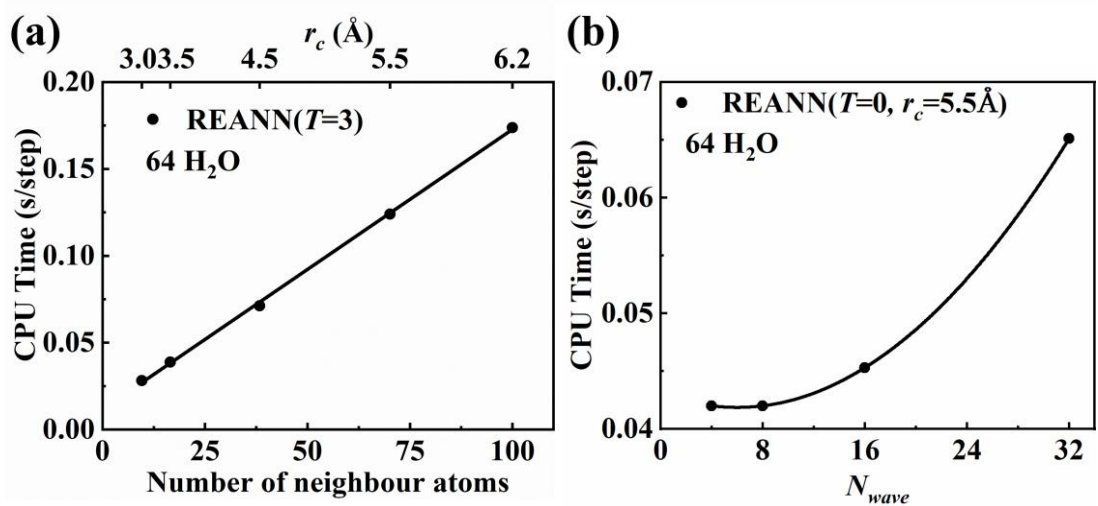

Fig. 12. (a) Computational costs (CPU time per MD step) of the REANN potentials as a function of (a) the number of neighbor atoms ($N_c$) and (b) the number of GTO ($n_{wave}$). All the results were tested via classical MD simulations of liquid water (300 K) including 64 $H_2O$ molecules using a single core of the Intel(R) Xeon 6132 2.60 GHz processor.

Table I. NN structures (represented by the number of neurons in each hidden layer), cutoff radii ($r_c$), and number of residual blocks (*nblock*) used in the training processes.

| System | NN structure[a] | *nblock*[a] | $r_c$ | $L$ | $n_{wave}$ | NN structure[b] | *nblock*[b] |
|---|---|---|---|---|---|---|---|
| MD17 | 128×128 | 1 | 8.0 | 2 | 20 | 128×128 | 1 |
| $CH_4$ | 128×128 | 2 | 8.0 | 2 | 20 | 128×128 | 2 |
| $N_4$ | 128×128 | 2 | 60.0 | 2 | 15 | 128×128 | 2 |
| $CO_2$+Ni(100) | 64×64 | 1 | 6.0 | 2 | 7 | 64×64 | 1 |
| Water-I[c] | 32×16 | 1 | 3.0~6.2 | 2 | 7 | 16×16 | 1 |
| $SiO_2$ (EANN) | 32×32 | 1 | 5.0 | 2 | 5 | / | / |
| $SiO_2$ (REANN) | 32×32 | 1 | 5.0 | 2 | 5 | 32×32 | 1 |
| $H_2O$ | 128×128 | 1 | 4.0 | 2 | 7 | / | / |
| $(H_2O)_2$ | 128×128 | 1 | 7.0 | 2 | 7 | 128×128 | 1 |
| $(H_5O_2)^+$ | 128×128 | 1 | 7.0 | 2 | 7 | 128×128 | 1 |
| Water-II[d] | 128×128 | 1 | 5.0 | 2 | 7 | 128×128 | 1 |

[a] and [b] represent the atomic NN structure for atomic energies and orbital coefficients, respectively. The NN structure is given as defined in "input_nn". In the realistic REANN model, a layer has been appended to the NN structure for the convenience of constructing residual NN using shortcut connections.

[c]Water-I refers to the dataset reported in Ref. [113].

[d]Water-II refers to the dataset reported in Ref. [56].

Table II. Test MAEs for MD17 with 1000 training data points (energies and atomic forces are given in meV and meV/Å, as the first and second numbers in each column). Best results are marked in bold.

| System | sGDML[32] | FCHL[105] | GM-sNN[10] | EANN[8] | SchNet[6] | PhysNet[9] | REANN |
|---|---|---|---|---|---|---|---|
| Aspirin | 8.2, 29.5 | 7.9, 20.7 | 16.5, 29.9 | 14.1, 43.0 | 16.0, 58.5 | 10.0, 26.2 | **6.7**, **14.6** |
| Ethanol | 3.0, 14.3 | **2.3**, **5.9** | 4.3, 14.3 | 4.4, 20.3 | 3.5, 16.9 | 2.6, 6.9 | 2.4, 6.6 |
| Malonaldehyde | 4.3, 17.8 | 3.5, 10.6 | 5.2, 19.5 | 5.9, 26.9 | 5.6, 28.6 | 4.1, 13.8 | **3.4**, **9.2** |
| Naphthalene | 5.2, 4.8 | 5.1, 6.5 | 7.4, 15.6 | **5.0**, 11.5 | 6.9, 25.2 | 6.2, 13.4 | **5.0**, **3.6** |
| Salicylic acid | 5.2, 12.1 | **4.9**, 9.6 | 8.2, 21.2 | 6.1, 22.3 | 8.7, 36.9 | 5.5, 14.6 | **4.9**, **9.0** |
| Toluene | 4.3, 6.1 | **4.2**, 8.8 | 6.5, 14.7 | 4.8, 16.6 | 5.2, 24.7 | 4.3, 8.3 | 4.1, **4.2** |
| Uracil | 4.8, 10.4 | **4.5**, **4.6** | 5.2, 14.3 | 4.8, 15.3 | 6.1, 24.3 | 4.7, 9.5 | 4.7, 7.4 |

Table III. The same as Table II but with 50000 training data points.

| System | DPMD[104] | GM-sNN[10] | GM-dNN[10] | EANN[8] | SchNet[6] | PhysNet[9] | REANN |
|---|---|---|---|---|---|---|---|
| Aspirin | 6.7, 12.1 | 8.2, 11.3 | **3.0**, 6.1 | 6.5, 18.9 | 5.2, 14.3 | 5.2, **2.6** | 5.3, 3.4 |
| Ethanol | 2.2, 3.1 | 2.2, 2.6 | 5.6, 5.2 | **2.1**, 4.0 | 2.2, 2.2 | 2.2, 1.3 | 2.2, **1.1** |
| Malonaldehyde | 3.3, 4.4 | 3.0, 3.5 | **2.2**, 1.7 | 3.1, 5.1 | 3.5, 3.5 | 3.0, 1.7 | 3.1, **1.4** |
| Naphthalene | 5.2, 5.5 | 5.6, 5.6 | **3.0**, 2.2 | 4.4, 8.8 | 4.8, 4.8 | 5.2, **1.7** | 4.9, 2.5 |
| Salicylic acid | 5.0, 6.6 | 4.8, 6.1 | 4.8, 3.5 | 4.5, 8.2 | **4.3**, 8.2 | 4.8, **1.7** | 4.6, 1.9 |
| Toluene | 4.4, 5.8 | 4.3, 4.3 | 4.8, 3.5 | 4.3, 9.1 | **3.9**, 3.9 | 4.3, **1.3** | 4.0, 1.6 |
| Uracil | 4.7, 2.8 | 4.3, 3.0 | **3.9**, 2.6 | 4.3, 3.6 | 4.3, 4.8 | 4.3, **1.3** | 4.5, 2.1 |

Table IV. Comparison of the test RMSEs of various ML models in energies (meV) and forces (meV/Å) for reactive systems, including the $N_4$ dataset in gas phase[108-110] (upper, energy only) and the $CO_2$+Ni(100)[111] dataset in gas-surface system (lower).

| | Model | PIP[108] | MEG1[109] | MEG2[110] | MB-PIP-NN[110] | REANN |
|---|---|---|---|---|---|---|
| $N_4$ | Energy | 7000 | 4683 | 420 | 120 | 65 |
| $CO_2$+Ni(100) | Model | BPNN[111] | | | REANN | |
| | Energy | 15 | | | 8.5 | |
| | Force | 26 | | | 12 | |

The PIP[108] and MEG1[109] (mixed-exponential Gaussians) PES was fitted to an incomplete data set with 16435 and 16534 configurations respectively, but the error was evaluated over all 21406 configurations as reported in Ref. [110].

Table V. Comparison of test MAEs of the EANN and REANN models in this work for the bulk $SiO_2$ with 60 atoms in energies (meV/atom) and forces (meV/Å), and previous NN models with descriptors ACSF (atom centered symmetry function), wACSF (weighted ACSF), EAD, SO3 (power spectrum), and SO4 (bispectrum) models reported in Ref. [80]. Note that these errors of reported in Ref. 79 were evaluated on 3840 points only.

| Model | ACSF | wACSF | EAD | SO3 | SO4 | EANN | REANN |
|---|---|---|---|---|---|---|---|
| Energy | 1.3 | 2.1 | 4.8 | 1.4 | 3.3 | 1.1 | 0.6 |
| Force | 81.2 | 141.8 | 259.0 | 115.1 | 204.2 | 69.9 | 32.2 |

Table VI. Comparison of relative RMSEs for permanent dipole moments and polarizability tensors by the SA-GAP, EANN and REANN models of several water-related system.

| System | μ | | | α | | |
|---|---|---|---|---|---|---|
| | SA-GAP* | EANN | REANN | SA-GAP* | EANN | REANN |
| $H_2O$ | ~0.11% | **0.02%** | 0.05% | ~0.02%/0.12% | **0.02%** | 0.06% |
| $(H_2O)_2$ | ~5.3% | 6.6% | 3.0% | ~6.4%/7.8% | 4.2% | 1.6% |
| $(H_5O_2)^+$ | ~2.4% | 1.3% | **0.4%** | ~3.8%/0.97% | 0.3% | 0.1% |
| Liquid water | \ | 16% | **15%** | ~5.8%/19%** | 2.2% | 2.1% |

* The SA-GAP model[56] is constructed in the irreducible spherical tensor representation, where the isotropic and anisotropic terms are learned separately and their errors are separated by "/".

**Values were reported for dielectric response tensors by indirect learning of molecular polarizability[56], which are used here for qualitative comparison only.